\documentclass[a4paper]{jpconf}

\usepackage{graphicx}
\usepackage{amssymb}

\begin{document}
\title{Regularized pseudopotential for mean-field calculations}

\author{K~Bennaceur$^1$, J~Dobaczewski$^{2,3,4}$, T~Haverinen$^{3,5}$ and M~Kortelainen$^{3,5}$}
\address{$^1$ Univ Lyon, Universit\'e Claude Bernard Lyon 1, CNRS, IPNL, UMR 5822,
4 rue E. Fermi, F-69622 Villeurbanne Cedex, France}
\address{$^2$ Department of Physics, University of York, Heslington, York YO10 5DD, United Kingdom}
\address{$^3$ Helsinki Institute of Physics, P.O. Box 64, 00014 University of Helsinki, Finland}
\address{$^4$ Institute of Theoretical Physics, Faculty of Physics, University of Warsaw,
Pasteura 5, 02-093 Warszawa, Poland}
\address{$^5$ Department of Physics, University of Jyv\"askyl\"a,
P.O. Box 35 (YFL), 40014 University of Jyv\"askyl\"a, Finland}
\ead{bennaceur@ipnl.in2p3.fr}

\begin{abstract}
We present preliminary results obtained with a finite-range two-body pseudopotential
complemented with zero-range spin-orbit and density-dependent terms. After discussing the penalty
function used to adjust parameters, we discuss predictions for binding energies of
spherical nuclei calculated at the mean-field level, and we compare them with those obtained
using the standard Gogny D1S finite-range effective interaction.
\end{abstract}

\section{Introduction}

A new class of pseudopotentials for nuclear structure were introduced several years
ago~\cite{Dobaczewski_2012,Raimondi_2014,Bennaceur_2017}. These pseudopotentials allow for a consistent formulation
of the low-energy energy-density-functional (EDF) approach in terms of effective theory.
Specifically, this can be done by considering a zero-range effective interaction
with derivative terms up to a given order $p=2n$, hereafter denoted N$^n$LO~\cite{carlsson},
and replacing the contact Dirac delta function by a regulator,
\begin{equation}
g_a(\mathbf r)=\frac{\mathrm e^{-\frac{\mathbf r^2}{a^2}}}{\left(a\sqrt{\pi}\right)^3}   ,
\end{equation}
where $a$ is the range of the obtained pseudopotential or regularization scale.

In this work we complemented the regularized pseudopotential
with the standard zero-range spin-orbit term
and two-body zero-range density-dependent effective
interaction. Therefore, the obtained EDF is meant to be used at the mean-field (single-reference) level.
The density-dependent term represents a convenient way to adjust the nucleon effective
mass in infinite nuclear matter to any reasonable value in the interval $0.70\lesssim m^*/m
\lesssim0.90$~\cite{ms}. For this zero-range density dependent term, we use the same
form as in the Gogny D1S interaction~\cite{d1s}, {\em i.e.},
\begin{equation}
\frac{1}{6}\,t_3\left(1+x_3\,\hat P_\sigma\right)\rho_0^{1/3}(\mathbf r_1)
\delta(\mathbf r_1-\mathbf r_2) ,
\end{equation}
where $\hat P_\sigma$ is the spin-exchange operator and $x_3$ is fixed to 1, so for time-even invariant
states,
this term does not contribute to pairing.
Finally, because of the zero-range nature of the spin-orbit term, we omitted
its contribution to the pairing channel.

The general EDF derived from this
pseudopotential~\cite{Bennaceur_2017}, including its particle-hole
and particle-particle parts, were limited to the local part. This
could constitute a significant restriction to its flexibility.
However, such a limitation reduces the number of free parameters to
be adjusted and simplifies implementations in the existing codes.

After presenting the ingredients of the penalty function used to adjust the
parameters, we present results obtained for binding energy of spherical nuclei
along with their comparison with those obtained for the Gogny D1S functional.

\section{Adjustments of parameters}

The pseudopotentials considered here contain 10 parameters at NLO, 14 at N$^2$LO
and 18 at N$^3$LO. We adjusted 15 series of parameters with effective masses of
0.70, 0.75, 0.80, 0.85, and 0.90 at NLO, N$^2$LO, and N$^3$LO. For each series, the
range $a$ of the regulator was varied from 0.8~fm to 1.6~fm.

The use of a penalty function containing data for finite
nuclei would not be sufficient to efficiently constrain these parameters
or even to constrain them at all. Typical reasons for this difficulty
are: appearance of finite-size instabilities, phase transitions to unphysical
states (for example, those characterized by a very large vector pairing) or numerical problems
related to compensations of large coupling constants with opposite signs.
To avoid these unwanted situations, the penalty function must contain
specially designed constrains that we list here, along with the nuclear data
and pseudo-data:
\begin{enumerate}
\item Empirical quantities in infinite nuclear matter: saturation density
$\rho_\mathrm{sat}$, binding energy per nucleon
in symmetric matter $E/A$, compression modulus
$K_\infty$, isoscalar effective mass $m^*/m$ symmetry
energy coefficient $J$, and its slope $L$, see Table~\ref{tab:inm}.
\item Decomposition of the potential energy in the different $(S,T)$
channel~\cite{Bal97,baldo} and binding energy per nucleon in neutron and polarized
matter.
\item Average pairing gap in infinite nuclear matter for $k_F=0.4$, $0.8$
and $1.2$~fm$^{-1}$ with the values obtained with D1S as targets.
\item Binding energies of the following 17 spherical (or approximated as
spherical) nuclei $^{36}$Ca, $^{40}$Ca, $^{48}$Ca, $^{54}$Ca,
$^{54}$Ni, $^{56}$Ni, $^{72}$Ni, $^{80}$Zr, $^{90}$Zr, $^{112}$Zr,
$^{100}$Sn, $^{132}$Sn, $^{138}$Sn, $^{178}$Pb,  $^{208}$Pb,  $^{214}$Pb,
and $^{216}$Th with a tolerance of 1~MeV if the binding energy is known
from experiment and 2~MeV if it is extrapolated (values are taken
from~\cite{Wang_2017}).
\item Proton density rms radii (taken from~\cite{unedf})
for $^{40}$Ca, $^{48}$Ca and $^{208}$Pb
with a tolerance of 0.02~fm and 0.03~fm for the one of $^{56}$Ni (extrapolated from
systematics);
\item Isovector density at the center of $^{208}$Pb and isoscalar density at the
center of $^{40}$Ca to avoid finite-size scalar-isovector ({\em i.e.} $S=0$, $T=1$)
instabilities. The use of the linear response (as in Ref.~\cite{hellemans}
for zero-range interactions) would lead to too much time-consuming calculations.
Therefore we use these two empirical constraints on these densities which are
observed to grow when a scalar-isovector instability
tends to develop. Possible instabilities in the vector channels ($S=1$) are not under
control in this series of fits.
\item Coupling constants for the vector pairing (given by eq.~(36)
in~\cite{Bennaceur_2017})
are constrained to be equal to $0\pm 5~\mathrm{MeV}\,\mathrm{fm}^3$ to avoid
transitions to unphysical states with unrealistically large vector pairing.
\end{enumerate}

These adjustments were performed in three steps:
\begin{enumerate}
\item First, we made exploratory adjustments (with fixed values for the effective
mass) trying to determine whether the other canonical
values for infinite nuclear matter were attainable and, in this case, what would be
their optimal values in average. We obtained $\rho_\mathrm{sat}=0.158~\mathrm{fm}^{-3}$,
for the saturation density, $J=29$~MeV for the symmetry energy coefficient and $L=15$~MeV
for its slope. This value for $L$ is very low compared with what is considered
as realistic
nowadays~\cite{li,Lattimer_2013,rocamaza} but we observed that larger values inevitably lead to finite-size
instabilities.
\item With effective mass and $\rho_\mathrm{sat}$,
$J$, and $L$ fixed to these values, and for each value of the effective mass and
order of the interaction, we systematically determined the ranges $a$ of the regulator
that give the lowest values of the penalty function.
\item With these values for the ranges fixed, we readjusted the parameters by relaxing
values of $\rho_\mathrm{sat}$, $m^*/m$, $J$, and $L$ and allowing for them
narrow tolerances of $\rho_\mathrm{sat}=0.158\pm0.003~\mathrm{fm}^{-3}$, $m^*/m=0.700\pm0.001$,
$J=29.0\pm0.5$~MeV, and $L=15.00\pm0.05$~MeV.
\end{enumerate}
The summary of targeted values and tolerances for infinite nuclear matter properties
are given in Table~\ref{tab:inm}.
\begin{table}
\caption{Infinite nuclear matter targeted properties and tolerances used for the final
step of the parameters adjustment.\label{tab:inm}}
\begin{center}
\begin{tabular}{lrrrrrr}
\br
Quantity  & $E/A$~[MeV] &  $\rho_\mathrm{sat}~[\mathrm{fm}]^{-3}$  & $K_\infty$~[MeV] & $ m^*/m$ & $J$~[MeV] & $L$~[MeV] \\
\br
Value     & -16.0 & 0.158               & 230       & 0.70-0.90 & 29.0 & 15.00 \\
\mr
Tolerance & 0.3   & 0.003               &   5       & 0.001     & 0.5 & 0.05 \\
\br
\end{tabular}
\end{center}
\end{table}
The targeted values and tolerances for all other data and pseudo-data will be
given and motivated with more details in a forthcoming article~\cite{future}.

\section{Results and discussion}

Questions concerning the dependence of the penalty function and observables on
the range of the regulator, covariance analysis of the parameters and propagation
of statistical errors on calculated quantities will not be discussed in this
contribution where we only report results for spherical nuclei. The sets of
parameters obtained by minimizing the penalty function will be given
in~\cite{future}.

We have built a set of 214 nuclei with even numbers of protons and neutrons which, according to the
predictions obtained with the Gogny D1S interaction~\cite{amedee}, can be considered as spherical
or almost spherical. In Table~\ref{tab:chi2}, we report the obtained average root mean squared deviations
$\sqrt{\overline{\Delta E^2}}$ and average deviations $\overline{\Delta E}$. We use the
subscript ``all'' when these quantities are calculated for the full set of 214 even-even nuclei and the
subscript ``fit'' when they are calculated for the 17 nuclei used in the penalty function only.

Since binding energies of nuclei are not the only ingredients used in the penalty function,
there is no reason for $\sqrt{\overline{\Delta E^2}}_\mathrm{fit}$ to decrease when more
parameters are used, {\em i.e.} to decrease with $n$ for interactions at N$^n$LO. Nonetheless,
we observe that it decreases with $n$ for all constrained values of the effective mass.
Interestingly, $\sqrt{\overline{\Delta E^2}}_\mathrm{all}$ is also a decreasing function
of $n$ for all values of the effective mass but for 0.7. This means that for $m^*/m\geqslant 0.75$,
the increase of the number of parameters in the pseudopotential improves its predictive
power, at least for the binding energies of spherical nuclei. The average deviation
$\overline{\Delta E}_\mathrm{fit}$ is also a decreasing function of $n$ while, in general,
$\overline{\Delta E}_\mathrm{all}$ has a less regular behaviour, although it does decrease
with $n$ for $m^*/m=0.85$.

\begin{table}
\caption{Average root mean squared deviation ($\sqrt{\Delta E^2}$) and average
deviation ($\overline{\Delta E}$) for 214 even-even nuclei (with subscript ``all'') and
for the 17 nuclei used in the penalty function (with subscript ``fit'') for the pseudopotentials
at NLO, N$^2$LO and N$^3$LO with effective mass from 0.70 to 0.90.\label{tab:chi2}}
\begin{center}
\begin{tabular}{lrrrrrr}
\br
\multicolumn{2}{c}{$m^*/m$} & \multicolumn{1}{c}{0.70} &
                              \multicolumn{1}{c}{0.75} &
                              \multicolumn{1}{c}{0.80} &
                              \multicolumn{1}{c}{0.85} &
                              \multicolumn{1}{c}{0.90} \\
\br
NLO$\!\!\!\!$ & $\sqrt{\overline{\Delta E^2}}_\mathrm{all}$ &
                                                 1.840 & 1.759 &  1.801 &  1.929 &  2.141 \\
    &$\overline{\Delta E}_\mathrm{all}$          & 0.382 & 0.029 & -0.301 & -0.633 & -0.950 \\
    &$\sqrt{\overline{\Delta E^2}}_\mathrm{fit}$ & 1.899 & 1.899 &  1.956 &  2.052 &  2.201 \\
    &$\overline{\Delta E}_\mathrm{fit}$          & 0.112 & 0.112 &  0.115 &  0.121 &  0.129 \\
\mr
N$^2$LO$\!\!\!\!$ & $\sqrt{\overline{\Delta E^2}}_\mathrm{all}$ &
                                                  2.028 & 1.827 &  1.709 &  1.594 &  1.540 \\
    &$\overline{\Delta E}_\mathrm{all}$          & 0.879 & 0.670 &  0.484 &  0.295 &  0.116 \\
    &$\sqrt{\overline{\Delta E^2}}_\mathrm{fit}$ & 1.893 & 1.741 &  1.690 &  1.610 &  1.602 \\
    &$\overline{\Delta E}_\mathrm{fit}$          & 0.111 & 0.102 &  0.099 &  0.095 &  0.094 \\
\mr
N$^3$LO$\!\!\!\!$ & $\sqrt{\overline{\Delta E^2}}_\mathrm{all}$ &
                                                  1.712 & 1.577 &  1.531 &  1.458 &  1.490 \\
    &$\overline{\Delta E}_\mathrm{all}$          & 0.378 & 0.231 & -0.048 & -0.105 & -0.313 \\
    &$\sqrt{\overline{\Delta E^2}}_\mathrm{fit}$ & 1.587 & 1.446 &  1.690 &  1.264 &  1.228 \\
    &$\overline{\Delta E}_\mathrm{fit}$          & 0.093 & 0.085 &  0.080 &  0.074 &  0.072 \\
\br
\end{tabular}
\end{center}
\end{table}

\begin{figure}
\caption{Binding energy residuals obtained with the pseudopotentials with $m^*/m$
constrained to 0.70 at order $n=1$, 2 and 3 (black dots) compared with the ones obtained
with the D1S Gogny interaction (open square).\label{fig:a}}
\begin{center}
\begin{tabular}{ccc}
\includegraphics[width=0.22\linewidth,angle=270,viewport=38 12 490 711,clip]{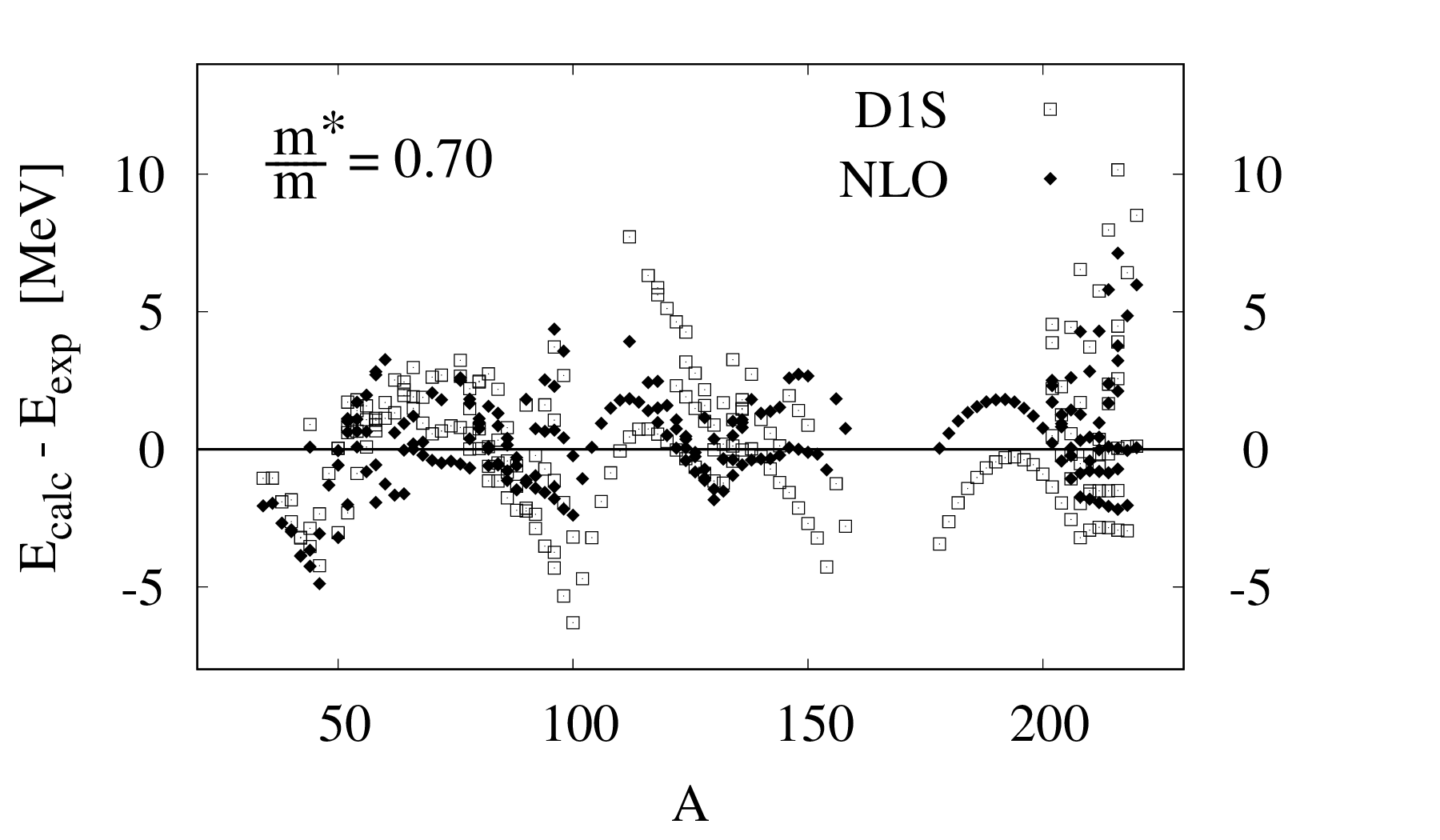} &
\includegraphics[width=0.22\linewidth,angle=270,viewport=38 118 490 711,clip]{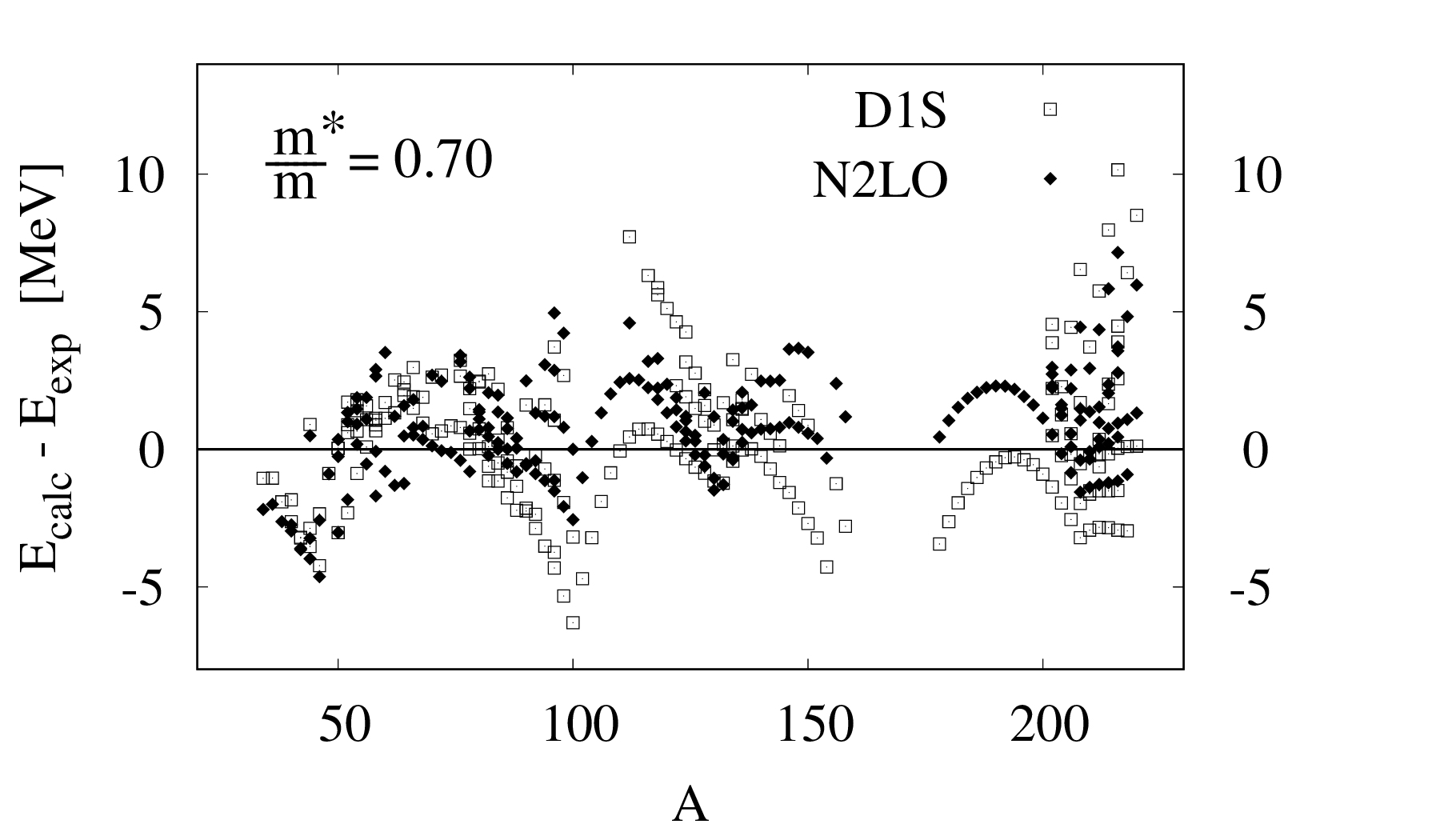} &
\includegraphics[width=0.22\linewidth,angle=270,viewport=38 118 490 711,clip]{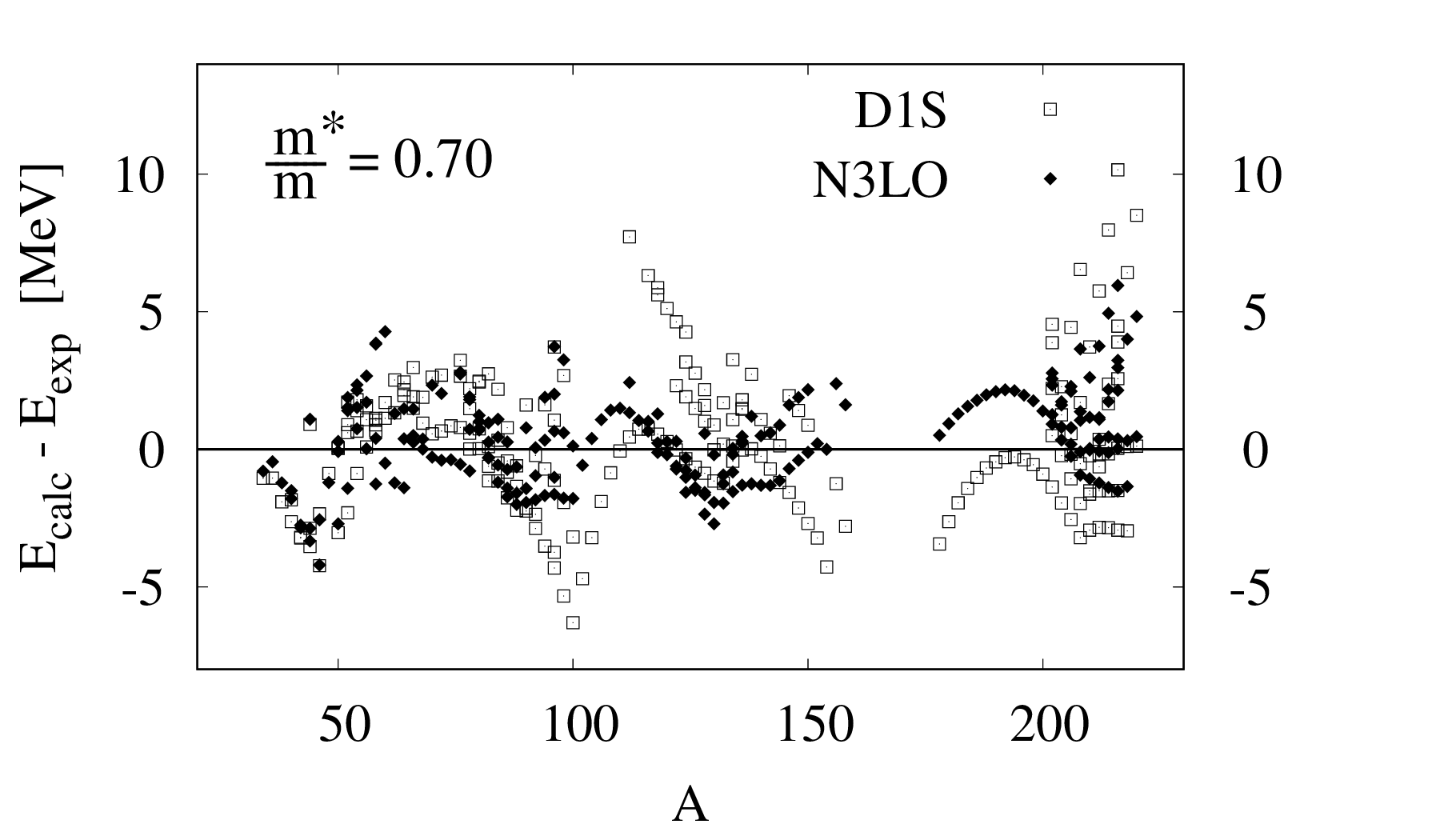}
\end{tabular}
\end{center}
\end{figure}

\begin{figure}
\caption{Same as figure~\ref{fig:a} for the pseudopotentials with $m^*/m$
constrained to 0.85.\label{fig:b}}
\begin{center}
\begin{tabular}{ccc}
\includegraphics[width=0.22\linewidth,angle=270,viewport=38 12 490 711,clip]{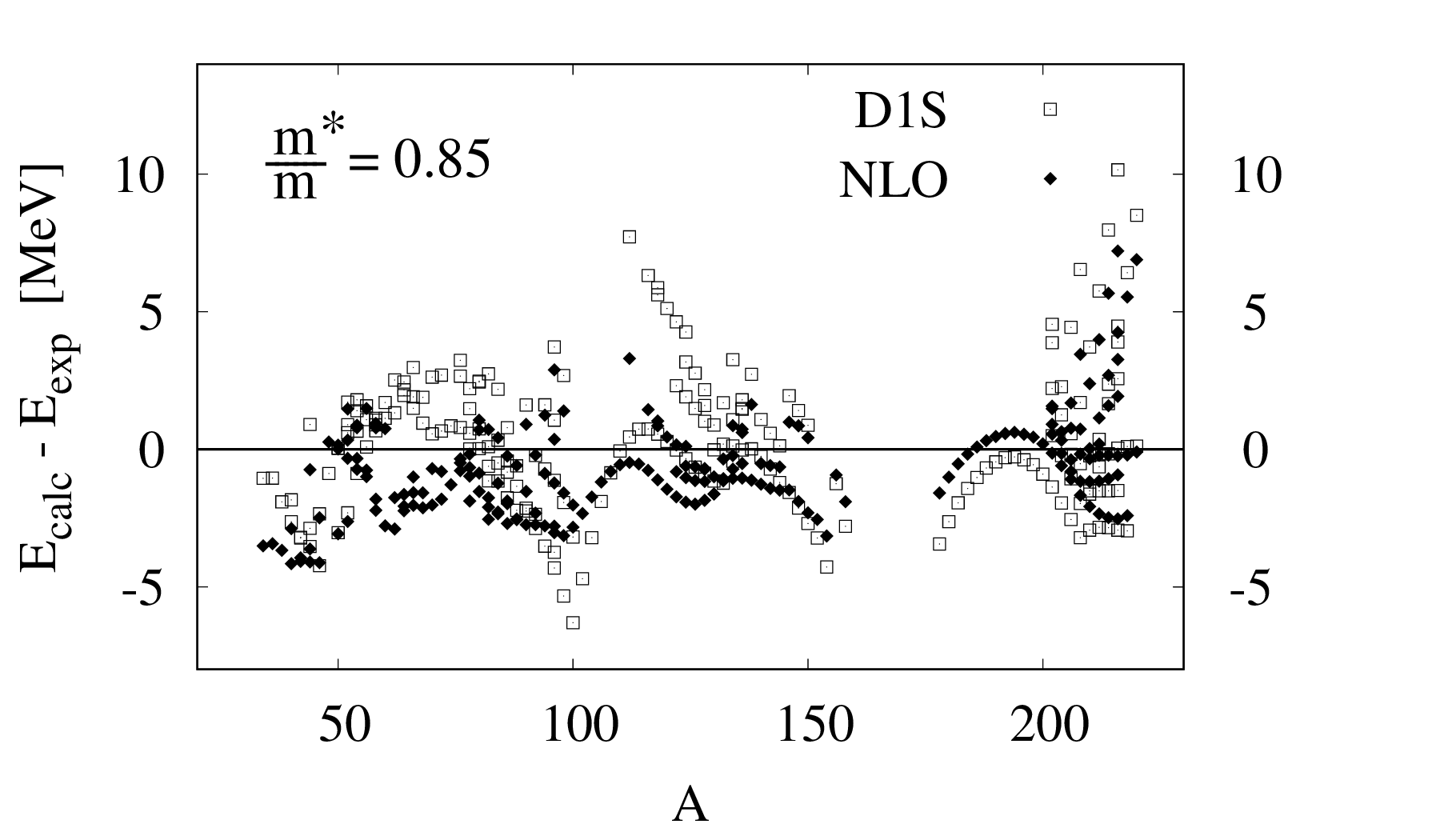} &
\includegraphics[width=0.22\linewidth,angle=270,viewport=38 118 490 711,clip]{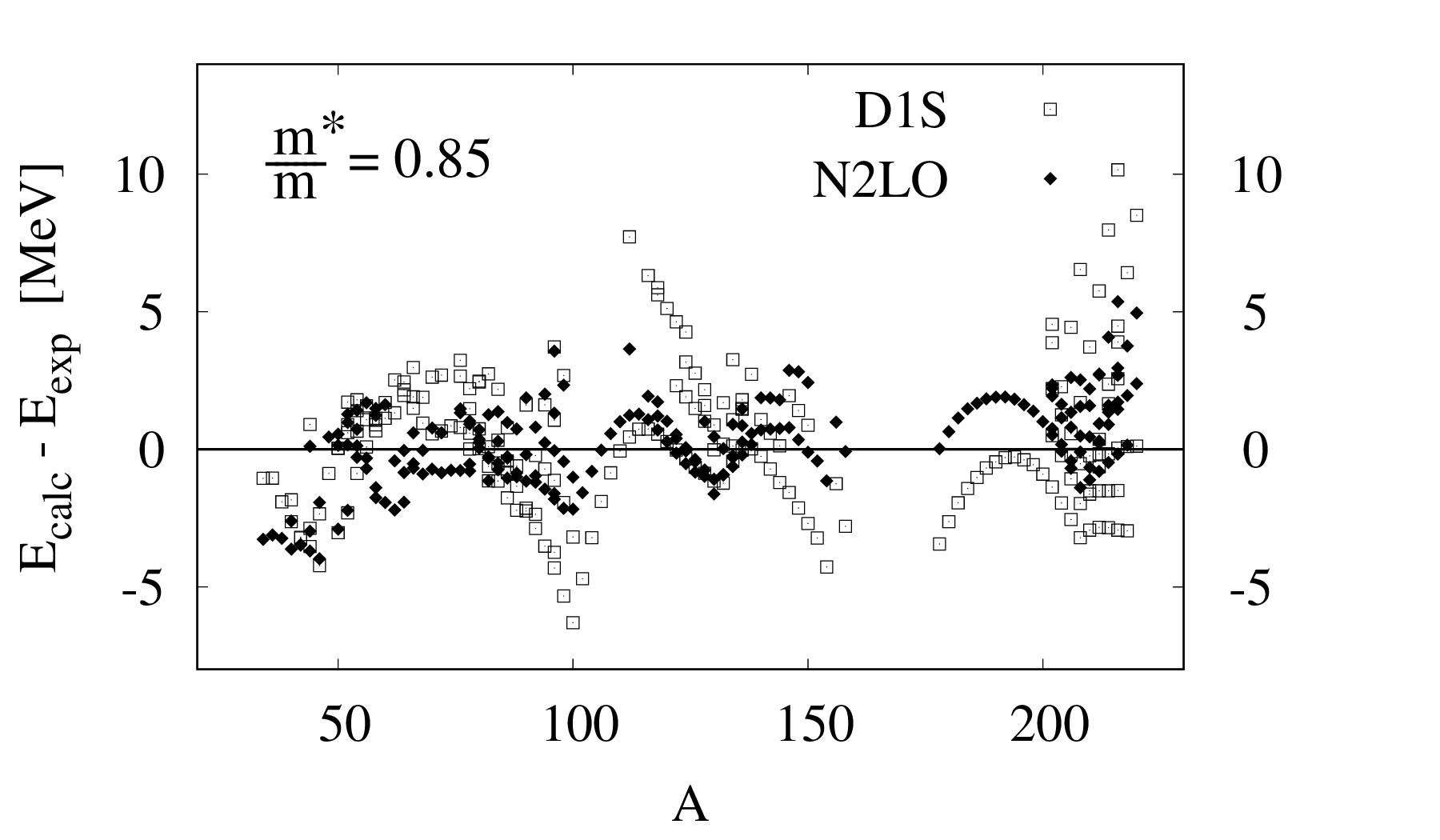} &
\includegraphics[width=0.22\linewidth,angle=270,viewport=38 118 490 711,clip]{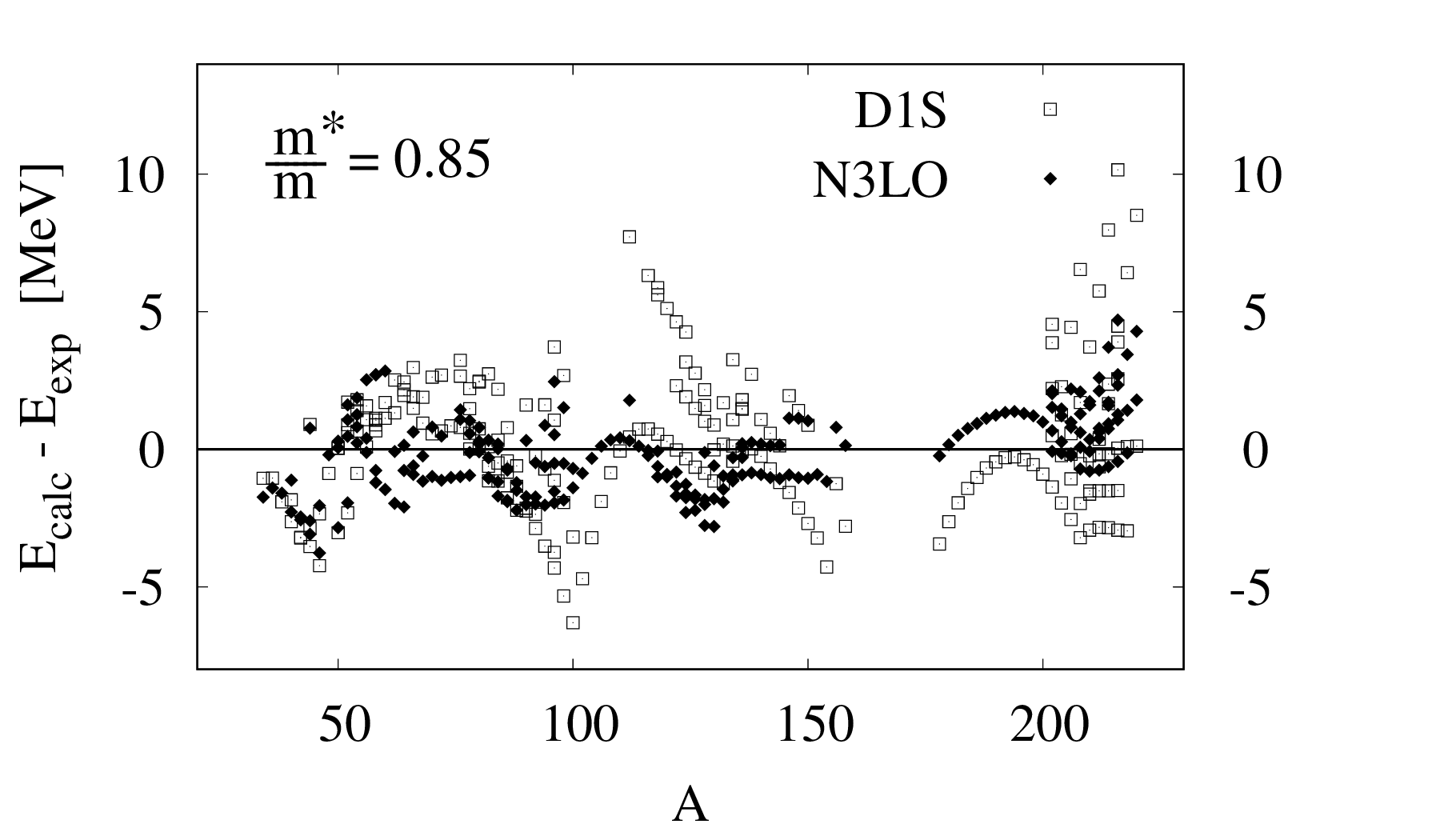}
\end{tabular}
\end{center}
\end{figure}

To visualize the global behaviour of the results obtained for the binding energies of
spherical nuclei, in Fig.~\ref{fig:a} we plotted the binding energy residuals obtained
for the set of 214 spherical nuclei and for the pseudopotential with $m^*/m=0.70$
at $n=1$, 2, 3. Similarly, in Fig.~\ref{fig:b}, we plotted those obtained for the pseudopotentials
with $m^*/m=0.85$. We chose these two values because, on the one hand, $m^*/m=0.70$
is close to the value obtained with D1S and, on the other hand, $m^*/m=0.85$ is the
effective mass that leads to the lowest value for $\sqrt{\overline{\Delta E^2}}_\mathrm{all}$
(obtained at N$^3$LO). In the same figures, to show the comparison with a commonly used
finite-range interaction, we also plotted the residuals obtained
for the Gogny D1S EDF. The comparison
should be considered with caution, because the Gogny D1S interaction, although often used at
the mean-field level only, is supposed to be used in beyond mean-field approaches, such
as the 5-dimensional Collective Hamiltonian (known as 5DCH~\cite{libert}) to provide
observables that can be compared with experimental data.

Both figures show that the residuals obtained for the regularized pseudopotentials
are more compressed around zero than those obtained for D1S.
Fig.~\ref{fig:a} explicitly exhibits the information already summarized in Table~\ref{tab:chi2}, {\em i.e.},
for the effective mass of 0.70, adding new parameters in the pseudopotential does not
significantly improve the predictive power for the binding energies of spherical nuclei.

Comparing the results shown in Figs.~\ref{fig:a} and~\ref{fig:b}, one can see that the typical
arches appearing in the residuals between shell closures are significantly
damped. Furthermore, one can see that this damping is more pronounced for higher order pseudopotentials.

\section{Conclusion}

In this article, we have reported results for binding energies of
spherical nuclei obtained for the new class of pseudopotentials introduced
several years ago~\cite{Dobaczewski_2012,Raimondi_2014,Bennaceur_2017}.
A more complete study including the discussion of proton radii, single particle energies, and properties
of deformed nuclei is in preparation. Although a definitive conclusion can only be drawn after a
comparison of a larger body of observables with data, the studied class of pseudopotential looks promising.
Possible improvements could still be the inclusion of non-local
terms and the use of regularized spin-orbit and tensor terms~\cite{Raimondi_2014,Bennaceur_2017},
which will be the subject of future developments.
Which part of correlations can be incorporated into the coupling constants of the
pseudopotential used at the mean-field (single-reference) level remains an open question.

\ack

This work was partially supported by the Academy of Finland under the Academy
project no. 318043, by the STFC Grants No.~ST/M006433/1 and
No.~ST/P003885/1, and by the Polish National Science Centre under Contract
No.~2018/31/B/ST2/02220.
We acknowledge the CSC-IT Center for Science Ltd. (Finland)
and the IN2P3 Computing Center (CNRS, Lyon-Villeurbanne, France)
for the allocation of computational resources.

\section*{References}

\bibliographystyle{iopart-num}

\bibliography{contribution}

\end{document}